\documentclass[nofootinbib,twocolumn,eqsecnum,floats,aps]{revtex4-1}
\usepackage{graphicx}
\usepackage{epsfig}
\usepackage{bm}
\newcommand{\sWA}{\mathsf{sWA}}
\newcommand{\T}{\mathsf{T}}

\newcommand{\tLN}{\mathsf{t}_{\Lambda N}}

\def\p {\mathsf{p}}

\def\bit{\begin{itemize}}
\def\eit{\end{itemize}}
\def\bnu{\begin{enumerate}}
\def\enu{\end{enumerate}}

\def\e {{\epsilon}}

\def\F {{{\cal F}}}

\def\O {{{\cal O}}}
\def\R {{{\cal R}}}

\def\nn{\nonumber }

\def\M {{{\cal M}}}
\def\x{\times}
\def\Ket#1{||#1 \rangle}
\def\Bra#1{\langle #1||}
\def\lsim{\:\raisebox{-0.5ex}{$\stackrel{\textstyle<}{\sim}$}\:}

\def\ie{{\em i.e., }}

\def\nn{\nonumber }
\def\be{\begin{equation}}
\def\ee{\end{equation}}
\def\br{\begin{eqnarray}}
\def\er{\end{eqnarray}}
\def\brn{\begin{eqnarray*}}
\def\ern{\end{eqnarray*}}
\def\etc{ {\it etc}}

\def\e {{\epsilon}}

\def\bra#1{\langle #1|}
\def\ket#1{|#1 \rangle}
\def\rf#1{{(\ref{#1})}}

\def\sixj#1#2#3#4#5#6{\left\{\negthinspace\begin{array}{ccc}
#1&#2&#3\\#4&#5&#6\end{array}\right\}}

\def\go{\rightarrow  }
\def\etal {\emph{et al.}}

\def\sqi{\frac{1}{\sqrt{2}}}
\def\fot{\frac{1}{2}}

\def\J {{{\cal J}}}

\include{def.tex}
\begin{document}
%
\title{ Relationships between nonmesonic-weak-decays in different hypernuclei}
\author{Franjo Krmpoti\'c}
\affiliation{Instituto de F\'isica La Plata, CONICET, 1900 La
Plata, Argentina, and Facultad de Ciencias Astron\'omicas y Geof\'isicas,
Universidad Nacional de La Plata, 1900 La Plata, Argentina.}

\date{\today}
\begin{abstract}
Using as a tool the $s$-wave approximation ($\sWA$), this work  demonstrates
 that the  nonmesonic weak decay transition rates $\Gamma_{n}$ and $\Gamma_{p}$ can
be expressed in all hypernuclei up to $^{29}_\Lambda $Si (and very
likely in heavier ones too) in the same way as in the $s$-shell
hypernuclei, \ie as a linear combination of only three elementary
transition rates. This finding leads to the analytic  prediction
that, independently of the transition mechanism, all  hypernuclei
that are on the stability line (${\sf N}={\sf Z}$),
 \ie $^5_\Lambda $He, $^7_\Lambda $Li, $^9_\Lambda $Be, $^{11}_\Lambda$B,  $^{13}_\Lambda $C,
 $^{17}_\Lambda $O, $^{29}_\Lambda $Si, \etc~ should  roughly  have the same
 ratio $\Gamma_{n}/\Gamma_{p}$, the magnitude of which rapidly increases when
 one approaches the neutron drip-line (${\sf N}\gg {\sf Z}$), and opposite
 happens
 when one goes toward the proton drip-line (${\sf N}\ll {\sf Z}$).

\end{abstract}

\maketitle
\section{Introduction}
 The nonmesonic weak decay   (NMWD)  of $\Lambda$ hypernuclei, $\Lambda
N\go nN$, takes place only within a nuclear  environment with the decay
rate $\Gamma_N$ ($N=p,n$). Without producing any additional
on-shell particle (as does  the mesonic weak decay  $\Lambda\go\pi N$),
 the mass is changed by $176$ MeV, and the strangeness by
 $|\Delta {\sf S}|=1$, which implies the
most radical modification  of an elementary particle within the
nucleus.  At the same time it offers the best opportunity to
study the strangeness-changing  interaction between hadrons,
 and is the main decay channel for medium and heavy hypernuclei.

With the incorporation of  strangeness, the radioactivity
domain is  extended to three dimensions $({\sf N,Z,S})$, which, because of the additional
binding due to the $\Lambda$-hyperon, is even richer
in elements than the ordinary $({\sf N,Z})$
domain.
  (For instance, while the one-neutron separation energy in $^{20}$C is $1.01$ MeV,
   it is $1.63$ MeV in $^{21}_\Lambda$C ~\cite{Sa08}.)
 This attribute of hypernuclei has motivated a recent proposal to  produce  neutron
rich $\Lambda$-hypernuclei at the Japan Proton Accelerator Research Complex (J-PARC),
including $^{9}_\Lambda$He~\cite{Sa09}
\footnote{ It is also speculated  that the NMWD
 could have an  important role in the stability of rotating neutron
  stars with respect to gravitational
wave emission~\cite{Da04, Ju08}.}.

Important  experimental efforts  have been invested in hypernuclear
weak physics during the last few years
\cite{Ki03,Ok04,Ok05,Ou05,Ka06,Ki06,Par07,Bh07,Ag08,Ag09,Ki09,Ag10}.
The correlative theoretical advances in our knowledge of the NMWD,
have been also quite significant
 \cite{Du96,Pa97,Ra97,Sa00,Al00,Ju01,Sa02,Pa02,It02,Ba02,Ba03,Kr03,Ga03,Ba04,Pa04,Ga04,Ga05,Sa05,Al05,Ba05,Ba07,Ch07,Bau07,It08,Ba08,Bau09,Co09,
 Bau09a,Bau09b,Bau09c,Bau10,Bau10a,Bau10c}.
 For recent review articles see Refs. \cite{Al02,Ou04,Pa07,Ch08}.
The ratio $\Gamma_{n/p}\equiv\Gamma_{n}/\Gamma_{p}$, together with
the asymmetry parameter ratio $a_\Lambda$ for  emission of
 protons from polarized hypernuclei~\cite{Ba07,Ch08}, has been in the past and
still  are the main concerns in the physics of NMWD. For a long
time the large experimental value for the $\Gamma_{n/p}$  ratio
(close to unity) remained unexplained. But, recent improved data
tend to converge to $\cong 0.5$~\cite{Ok05,Ou05,Ka06,Ki06}, both
for $^{5}_\Lambda$He ($s$-shell) and  $^{12}_\Lambda$C
($p$-shell), indicating similarity in the transition mechanism.

 In the meantime the theoretical estimates of $\Gamma_{n/p}$, done within  the
 one meson-xchange (OME) model, have increased. For instance, Parre\~{n}o, and
Ramos~\cite{Pa02} have found
$\Gamma_{n/p}(^{5}_\Lambda$He$)=0.34-0.46$, and
$\Gamma_{n/p}(^{12}_\Lambda$C$)=0.29-0.34$, when the  exchanges of
the complete pseudoscalar ($\pi, K, \eta$) and vector
($\rho,\omega,K^*$) meson octets are taken into account, with the
weak coupling constants obtained from soft meson theorems and
$SU(6)_W$~\cite{Pa97,Du96}. The dominant role is  played by the
exchange of  pion and kaon mesons, and when their effect is
combined with the Direct-Quark (DQ) model, the value of the
$n/p$ ratio is increased up to $0.70$~\cite{Sa00,Sa02}.
 However, these transition mechanisms
continue to predict too large and negative value for $a_\Lambda$.
There are two recent proposals to bring this value into agreement
with experiments by going beyond the OME+DQ models. The first
considers incorporating new scalar-isoscalar terms induced by
$2\pi$-exchanges~\cite{Ch07}.( See also Refs.~\cite{Ju01,Pa04} on
the relevance of these terms.) In the second, in addition to the
model of $\pi + 2\pi/\rho +2\pi/\sigma + 2p/s +\omega + K$
exchanges, is introduced the axial-vector $a_1$-meson exchange~\cite{It08}.

Quite recently we have discussed
the parameter $a_\Lambda$  within the independent particle shell model (IPSM), together with
 the $s$-wave approximation ($\sWA$)~\cite{Ba07}. The corollary of this study was that, independently of
 the NMWD dynamics,  this observable has the same value in all hypernuclei that have totally full proton subshells,
such as  $^{5}_\Lambda$He and  $^{12}_\Lambda$C, and very likely also in the remaining hypernuclei.
 This result is a direct consequence of the fact that
 $a_\Lambda$, same as $\Gamma_{n/p}$,   is  a ratio of two transition rates, which makes it,
in  absence of Final State Interactions (FSI),  dependent purely on the dynamical
features of the NMWD.

 The aim of this work is twofold. First, we establish the link between the theoretical
formalism for the NMWD of the $s$-shell hypernuclei originally introduced by Block
and Dalitz~\cite{Bl63}, and the general formalism used presently for any type of hypernuclei.
Second, we show that the IPSM framework, together with
 the $\sWA$, allow us to formulate the rates $\Gamma_N$ within
the $p,d,$\etc~ shells
 in terms of the  $s$-shell nuclear matrix elements (NME).
 Previous research in this direction has been done by Alberico, and
  Garbarino~\cite{Al00} and  by Cohen~\cite{Co90}.
Later on,
it is demonstrated that regardless  of the decay mechanism:  i) all hypernuclei
with the same number of protons and neutrons  (\ie with ${\sf Z}= {\sf N})$ should have the same  ratio
 $\Gamma_{n/p}$,  ii) the value of this observable increases (decreases)
 as the neutron (proton) excess is enlarged, and
 iii) simple analytic relationships exist between
   $\Gamma_{n}$,  $\Gamma_{p}$, and $\Gamma_{n/p}$ in  different hypernuclei
 with the same  mass number $A$.
The derivation of these results, same as those on the parameter $a_\Lambda$~\cite{Ba07},
is based on the assumption that  the emission of the nucleons $N$ from
different single-particle states is affected in a
similar way by the FSI.
Then, before presenting the formalism, it  might be convenient
to  comment on the  relationship between the ratio $\Gamma_n/\Gamma_p$
and the FSI.

The primary partial decay rates $\Gamma_N$  are in principle derivable from the
measurements of emitted
nucleons $n$ and $N$  spectra. These  are: i) the single-nucleon spectra  $S(E_N)$,
 as a function of one-nucleon kinetic energies $E_N$, and ii) $nN$ coincidence spectra
  $S(E_{nN})$,  and $S(\cos\theta_{nN})$, as  functions of
 the sum of  kinetic energies $E_{nN}=E_n+E_N$,
 and  the opening angle $\theta_{nN}$, respectively.
From these spectra are determined the  numbers of protons N$_p$,
and neutrons N$_n$, and numbers of pairs N$_{nn}$, and  N$_{np}$, which are not related
in a simple way with $\Gamma_n$ and $\Gamma_p$. This is because
not all  primary nucleons, originated by the NMWD, are measured.
In  propagating within the nuclear environment they interact with
the surrounding nucleons, and in some
  cases they change their momenta, and  energies,  some
of them  even can be absorbed by the medium, and  emission of
additional (secondary) nucleons  can take place  as
well~\cite{Bau09a,Bau09b,Bau09c,Bau10,Bau10a,Bau10c}.
All these processes represent a complicated many-body problem, and
are generically designated as
FSI. To describe them, keeping  the calculations feasible,
 are indispensable model assumptions, and the FSI are usually  simulated by a semi-classical
 model,  developed by Ramos \etal~\cite{Ra97}, and called the   intranuclear cascade (INC) model.
 This model interrelates the rates $\Gamma_n$, and $\Gamma_p$ with the numbers
N$_n$, N$_p$, N$_{nn}$, and N$_{np}$, and therefore,
 as stressed recently by Bauer and  Garbarino \cite{Bau10a}, the  FSI described by the INC model
  should not be included in the evaluation of decay rates
$\Gamma_n$, and $\Gamma_p$.
However, not all FSI are considered within the INC model, and which additional
FSI contribute to the NMWD spectra and
decay rates, and how and which of them should be included in the calculation
are nontrivial questions. Some candidates  are as follows:

(i) Short range correlations (SRCs) acting on
 final  $nN$ states; here one starts from the plane wave approximation for the outgoing nucleons
  and the SRCs are incorporated {\it a posteriori}, either
 phenomenologically  through Jastrow-like SRC functions,
  or by solving the Bethe-Goldstone equation. The first approach
 is used   within both nuclear matter
 ~\cite{Ba04,Bau09a,Bau09b,Bau09c}, and finite nuclei calculations
 ~\cite{Pa97,Ba02,Ba03,Kr03}, and the second one only
in the shell-model-type calculations
  ~\cite{It08,Pa97,Pa02,It02}.

(ii) Self-energy and vertex particle-hole corrections, and RPA-like rescattering
effects (see, for instance, the diagrams
(b)-(d) in Fig. 2 of Ref.~\cite{Ry94}).
It is not  known  whether these  FSI  contribute coherently or incoherently,
and it can even happen that (b) and (c)  cancel out, as do the  divergences in the vertex,
and fermion self-energy corrections in the QED, because of  the Ward identity.
 (Something similar happens also in the nuclear particle-phonon-coupling model.)
The first ones can be associated
with the mean-field effects on the single-particle wave functions engendered by  an
energy-dependent complex optical potential~\cite{Co09}.

(iii) Interactions of the deep-hole states (which become  highly excited
states in the continuum after the NMWD)  with  more complicated
configurations ($2h1p, 3h2p\cdots$, collective states,\etc), which spread their
transition  strengths in relatively large energy  intervals~\cite{Ba08}.

There is no theoretical study in the literature on the NMWD
encompassing  all aspects of the FSI. The development of a
microscopic many-body model for the FSI described by the INC model would
be also extremely welcome, and   so far only  in
Refs.~\cite{Bau07,Bau09c} were  the first steps taken toward this
goal. Finally, the two-body induced NMWD $\Lambda NN\go nNN$,
  which has been recently measured \cite{Ki09}, should be also considered
 \footnote{It also has been shown that the kinematical and
nonlocal  and kinematical effects on the NMWD could be sizable~\cite{Ba03}.}.
 Briefly, the issue of FSI  is a tough nut to crack,
 and  a lot of theoretical work has to be done still,
particularly in relation to the recently measured spectra $S(E_N)$, $S(E_{nN})$,
and $S(\cos\theta_{nN})$~\cite{Ki03,Ok04,Ok05,Ou05,Ka06,Ki06,Par07,Bh07,Ag08,Ag09,Ki09,Ag10},
which are certainly  affected by them.
 However, as the purpose  of the present
contribution is not to make progress in this direction,  among all possible
FSI, only  the SRCs will be considered here. This is done
 phenomenologically, and  initial $\Lambda N$ state SRCs
are included on  the same footing~\cite{Pa97,Ba02,Ba03,Kr03}. It
is our belief that this is a fair approximation for the
objectives of the present work.

\section{Decay Rates}

 To derive the NMWD rate within the IPSM we start with the Fermi Golden Rule.
For  a hypernucleus, in its ground state with spin
$J_I$ and energy $E_{J_I}$, decaying to: i)  several
states $\alpha_N$ in the residual nuclei with spins $J_F$ and energies $E_{\alpha_NJ_F}$,
and ii) two free nucleons $n$ and $N$, with  momenta $\bm{p}_n$, and $\bm{p}_N$,  kinetic
energies  $E_{n}=\bm{p}_n^2/2M$, and $E_{N}=\bm{p}_N^2/2M$,  and  total spin $S$,
reads~\cite{Ba02,Ba03,Kr03}
\begin{eqnarray}
\Gamma_N &=& 2\pi \sum_{S\alpha_NJ_F}
\int\delta(\Delta_{\alpha_NJ_F}-E_R-E_n-E_{N})
\nn\\
&\x& |\bra{\bm{p}_n\bm{p}_N S;\alpha_NJ_F}V\ket{J_I}|^2 \frac{d{\bf
p}_n}{(2\pi)^3}\frac{d{\bf p}_N}{(2\pi)^3}, \label{2.1}
\end{eqnarray}
where for sake of simplicity we have suppressed the magnetic
quantum numbers. The NMWD dynamics, contained within the  weak
hypernuclear transition potential $V$, will be described  by the
OME model. The wave functions for the kets $\ket{\bm{p}_n\bm{p}_N
SM_SJ_FM_F}$ and $\ket{J_IM_I}$ are assumed to be antisymmetrized
and normalized, and the two emitted nucleons $n$ and $N$ are
described by plane waves.  Initial and final SRCs are included
phenomenologically at a Jastrow-like level, while the finite
nucleon size effects at the interaction vertices are gauged by
monopole form factors~\cite{Pa97,Ba02}. Moreover,
 \be
 E_R =\frac{|\bm{p}_n+\bm{p}_N|^2}{2M(A-2)}= \frac{E_n+E_{N} + 2\cos\theta_{nN} \sqrt{E_n
E_N}}{A - 2},
 \label{2.2}\ee
 is the
recoil energy of the residual nucleus,  and
\begin{equation}
\Delta_{\alpha_NJ_F}=\Delta +E_{J_I}-E_{\alpha_NJ_F},
\label{2.3}
\end{equation}
with
$\Delta =M_\Lambda-M=176$ MeV
is the liberated energy.

It could be convenient to perform a transformation to the relative
and c.m. i) momenta: $\bm{p}=(\bm{p}_n-\bm{p}_N)/2 $,
$\bm{P}=\bm{p}_n+\bm{p}_N$, ii) coordinates
$\bm{r}=\bm{r}_n-\bm{r}_N$,  $\bm{R}=(\bm{r}_n+\bm{r}_N)/2$, and
iii) orbital angular momenta $\bm{l}$ and $\bm{L}$. The energy
conservation is  expressed as \be
E_n+E_{N}+E_R-\Delta_{\alpha_NJ_F}=\e_p+\e_P-\Delta_{\alpha_NJ_F}=0,
\label{2.4}\ee where \br \e_p&=&\frac{p^2}{M},~~~
E_R=\frac{P^2}{2M(A-2)},
~\nn\\
\e_P&=&\frac{P^2}{4M}\frac{A}{A-2}=\frac{A}{2}E_R, \label{2.5}\er
are, respectively, the energies of  the relative motion of the
outgoing pair, of the recoil, and of the total c.m. motion
(including the recoil).

Following  the analytical developments
done in Ref.~\cite{Ba02},   the transition rate can be expressed
as a function of the c.m. momentum $P$:
  \br
\Gamma_{N}&=&
\frac{2M}{\pi}\sqrt{\frac{A-2}{A}}\int dP
\sum_{\alpha_NJ_F}
\nn\\
&\x&P^2\sqrt{P^2_{\Delta_{\alpha_NJ_F}}-P^2}\mathcal{F}_{\alpha_NJ_F}(pP),
\label{2.6}\er
with
\begin{widetext}
\br
\mathcal{F}_{\alpha_NJ_F}(pP) &=&\hat{J}_I^{-2}\sum_{S\lambda lLTJ}
\left|\sum_{j_N} \M(plPL\lambda SJ\T;j_\Lambda j_N J\tLN)
\Bra{J_I}\left( a_{j_N}^\dag a_{j_\Lambda}^\dag\right)_{J}
\Ket{\alpha_NJ_F}\right|^2,
\nn\\
 \label{2.7}\er
\end{widetext}
where
\be
P_{\Delta_{\alpha_NJ_F}}=2\sqrt{\frac{A-2}{A}\Delta_{\alpha_NJ_F}},
 \label{2.8}\ee
 and
 \br
p&=&\fot \sqrt{\frac{A}{A-2}\left[P^2_{\Delta_{\alpha_NJ_F}}-P^2\right]},
\label{2.9}\er
It is clear that  the condition
$P\le P_{\Delta_{\alpha_NJ_F}}$ has to be fulfilled for each
final state $\ket{\alpha_NJ_F}$.
Moreover
\begin{eqnarray}
&&\M(plPL\lambda SJ\T;j_\Lambda j_N J\tLN)
=\sqi\left[1-(-)^{l+S+T}\right]
\nonumber \\ &\times&\O_L(P)
({lL\lambda SJ\T}|V(p)|{j_\Lambda j_N J\tLN}),
\label{2.10}
\end{eqnarray}
 where (and henceforth) the ket $|)$, unlike  $\ket{}$, indicates that the
state is not antisymmetrized,
 \be
 \O_L(P) =\int R^2dRj_L(PR){\rm R}_{0L}(b/\sqrt{2},R),
 \label{2.11}\ee
is the overlap of the c.m. radial wave functions   ${\rm R}_{0L}$,
and $j_L$ for the bound and outgoing particles, respectively, and
$b$ is the harmonic oscillator  size parameter.
More, $\bm{\lambda}=\bm{l}+\bm{L}$,
 $\T\equiv \{TM_T,M_T=m_{t_\Lambda}+m_{t_N}\}$, and $\tLN\equiv \{t_\Lambda=1/2,m_{t_\Lambda}=-1/2,
t_N=1/2,m_{t_N}\}$, with
$m_{t_p}=1/2$,  and $m_{t_n}=-1/2$, where   we have  assumed
that the $\Lambda N\go nN$ interaction occurs with the isospin
change $\Delta T=1/2$. Explicitly,
\be
|\tLN)=\left\{
\begin{array}{c}|T=1),~~~~~~~~~~~~~~~~~~~~~~~~\mbox{for}~~N=n\\
(|T=1)-|T=0))/\sqrt{2},~~\mbox{for}~~N=p\\
\end{array}\right..
\label{2.12}\ee It might be pertinent to mention that the factor
$(A-2)/A$ in Eqs. \rf{2.6}, \rf{2.8}, and \rf{2.9} comes from the
recoil effect,
 which, in the same way as
 the spreading of the deep hole states,
is relevant  for the NMWD spectra~\cite{Ba08,Bau09}, but its role is of minor importance for the
total transition rates $\Gamma_{N}$.

\subsection{Independent-article shell model}
 Up to now nothing has been  said about the initial state $\ket{J_I}$,  and
 final states $\ket{\alpha_NJ_F}$.
Within the IPSM  the following assumptions are made, which greatly
 simplify the numerical calculations:
 \\1) The initial hypernuclear state is taken as a $\Lambda$-particle in
single-particle state $j_\Lambda=0s_{1/2}$ weakly coupled to an
$(A-1)$ nuclear core of spin $J_C$, i.e.,
$\ket{J_I}\equiv\ket{(J_Cj_\Lambda)J_I}$.
\\2) When the nucleon inducing the decay  is the single-particle state
$j_N$ ($j\equiv nlj$), the final  residual nucleus states are:
$\ket{\alpha_N J_F}\equiv\ket{(J_Cj_N^{-1})J_F}$.
\\3. We adopt  the simplest version of the IPSM, in
which all the relevant particle states are assumed to be
stationary,  and the liberated energy is
\be
\Delta^j_{N} =\Delta  + \varepsilon_{\Lambda} +
\varepsilon^j_{N},
\label{2.13}\ee
where $N=p,n$,  and
the $\varepsilon$'s are single-particle energies.
 (The  non-stationary version of the IPSM is discussed in Ref.~\cite{Ba08}.)

Within this scheme, we get ~\cite{Ba02,Kr03}
 \br
\Gamma_{N}&=&\sum_{j}\Gamma_{N}^j;~~~\Gamma_{N}^j=\sum_{J=|j-1/2|}^{J=j+1/2}F^j_{NJ}\R^j_{NJ},
\label{2.14}\er
 where the summation goes over all single-particle
transition rates $\Gamma_{N}^j$, which in turn results from the sum
over the values of $\bm{J}=\bm{j}_N+\bm{j}_\Lambda$
 of products  of the spectroscopic factors $F^j_{NJ}$ with
the  partial $\Lambda n\go nN$ transition rates $\R_{NJ}^j$.
 For  the $s$-shell nuclei the later
have the same physical meaning as the quantities $R_{NJ}$ introduced
in the seminal work of Block and Dalitz~\cite{Bl63} (see also Ref.~\cite{Co90}),
 \ie $\R_{NJ}^{s_{1/2}}\equiv R_{NJ}$.
\footnote{In order to use here the same  notation for $R_{NJ}$ as in Ref.~\cite{Bl63},
as well as to write $\R_{NJ}^j$ instead of   $\R_{J}^{j_N}$, the   $j_N$
variable employed in previous publications is frequently split here in $j$, and   $N$.}

The spectroscopic factors $F^j_{NJ}$
are defined as
\br
F^j_{NJ}&=&\hat{J_I}^{-2}\sum_{J_F} |\Bra{J_I}\left( a_{j_N}^\dag
a_{j_\Lambda }^\dag\right)_{J}\Ket{J_F}|^2 \label{2.15}
\\
&=&\hat{J}^{2}\sum_{J_F}\sixj{J_C}{J_I}{j_\Lambda}{J}{j_N}{J_F}^2
|\Bra{J_C}a_{j_N}^\dag\Ket{J_F}|^2,
 \nn\er
with the notation $\hat{J}=\sqrt{2J+1}$, while the
 partial transition rates read
\br
\R_{NJ}^j&=&
\frac{2M_{\rm N}}{\pi}\sqrt{\frac{A}{A-2}}\int^{P_N^j }_0 dPP^2\sqrt{(P_N^j )^2-P^2}
\nn\\
&\x&\sum_{SlL\lambda T}|\M(plPL\lambda SJ\T;{j_\Lambda j_N J\tLN})|^2,
\label{2.16}\er
with
\be
{P}_N^j=2\sqrt{\frac{A-2}{A}M_{\rm N}\Delta_N^j}
 \label{2.17}\ee
the  maximum value of $P$ for each $j_N$, and
 \br
p&=&\fot \sqrt{\frac{A}{A-2}\left[(P_N^j )^2-P^2\right]},
\label{2.18}\er
the corresponding relative momentum.

It should be stressed that the most important virtue of the IPSM is
that the index $\alpha$ becomes superfluous,  and  the summation
on the final spins $J_F$ can be carried out without knowing the nuclear structure
of the initial and final nuclear states. This simplifies enormously the numerical calculations.
As far as we know,  the IPSM has been used to a great extent in all previous finite nucleus evaluation of the NMWD.

\subsection{$s$-wave approximation}

Gale\~ao~\cite{Gal04} has shown that the matrix elements in \rf{2.10} can be cast in the form
\br
({lL\lambda SJ\T}|V(p)|{j_\Lambda j_N J\tLN})
&=&\sum_{KS'{\sf l}}(lSK\T|V(p)|{\sf l}S'K\tLN)\nonumber \\ &\times&
C_{\sf l}(lL\lambda SJl_Nj_N;KS'),
\nonumber \\
\label{2.19}\er
with
 \br
&&C_{\sf l}(lL\lambda SJl_Nj_N;KS')
=(-)^{j_N+\fot+S+\lambda}\nonumber \\ &\times&
\hat{l}_N\hat{\lambda}\hat{j}_N \hat{S'}\hat{K}^2(0{\sf l}0Ll_N|000l_N l_N)
\sixj{J}{j_N}{\fot}{\fot}{S'}{l_N}
\nonumber\\ &\times&\sixj{L}{l}{\lambda}{S}{J}{K}\sixj{{ L}}{K}{J}{S'}{l_N}{{\sf l}},
\label{2.20}\er
where $(0\cdots|\cdots l_N)$ are the Moshinsky brackets \cite{Mo59}.
The plain $\sWA$ implies that we  make ${\sf l}= 0$ in \rf{2.19}, and \rf{2.20}, which leads to
\br
({lL\lambda SJ\T}|V(p)|{j_\Lambda j_N J\tLN})
&=&\sum_{K=0,1}C_0(l\lambda SJl_Nj_N;K),
\nonumber \\ &\times&
(lSK\T|V(p)|0KK\tLN),\nonumber \\
\label{2.21}\er
with
 \br
C_0(l\lambda SJl_Nj_N;K)
&=&\hat{K}^2\hat{\lambda}\hat{j}_N (000l_Nl_N|000l_N l_N)
\nn\\
&\x&\delta_{L l_N}(-)^{j_N+\fot+S+\lambda+l_N+K+J}
\nn\\
&\x&\sixj{J}{j_N}{\fot}{\fot}{K}{l_N}\sixj{K}{l}{S}{\lambda}{J}{l_N}.
\nn\\
\label{2.22}\er
In particular, for the $s$-shell  hypernuclei
 \be
C_0(ll SJ0,1/2;K)=\delta_{JK},
 \label{2.23}\ee
and using Eqs. \rf{2.10}, and \rf{2.12},  Eq. \rf{2.16} becomes
\br
&&\R_{NJ}^{s_ {1/2}}\equiv R_{NJ}=(1+\delta_{nN})
\frac{2M_N}{\pi}\sqrt{\frac{A}{A-2}}\nn\\
&\x&\int^{P_N }_0 dPP^2\sqrt{P_N^2-P^2}\O_0^2(P)
\label{2.24}\\
&\x&\sum_{SlT}\left[1-(-)^{l+S+T}\right]|(lSJ\T|V(p)|0JJ\T)|^2,
\nn\er
where $P_{N}\equiv P^{s_{1/2}}_{N}$, and
\br
\O_0^2(P)&=&\left(\frac{\pi}{2}\right)^{1/2}b^{3}e^{-(Pb)^2/2}.
\label{2.25}\er

As it is well known, the corresponding transition rates
\br
\Gamma_{N}(^3_\Lambda{\rm H})&=&\frac{3}{4}R_{N0}+\frac{1}{4}R_{N1},
\nn\\
\Gamma_{n}(^4_\Lambda{\rm H})&=&\frac{1}{2}R_{n0}+\frac{3}{2}R_{n1},
\hspace{0.5cm}
\Gamma_{p}(^4_\Lambda{\rm H})=R_{p0},
\nn\\
\Gamma_{p}(^4_\Lambda{\rm He})&=&\frac{1}{2}R_{p0}+\frac{3}{2}R_{p1},
\hspace{0.5cm}
\Gamma_{n}(^4_\Lambda{\rm He})=R_{n0},
\nn\\
\Gamma_{N}(^5_\Lambda{\rm He})&=&\frac{1}{2}R_{N0}+\frac{3}{2}R_{N1},
\label{2.26}\er
with $N=n,p$, depend only on four single-particle transition rates
$R_{n0}$, $R_{n1}$, $R_{p0}$ and $R_{p1}$.


Here we will express the $\Gamma_{N}$'s of heavier hypernuclei
in the same way as was done in Eq. \rf{2.26} for the $s$-shell hypernuclei,
\ie as a linear combination of $R_{N0}$ and $R_{N1}$ only:
\br
\Gamma_{N}&=&\F_{N0}R_{N0}+\F_{N1}R_{N1}.
\label{2.27}\er
To derive  the generalized spectroscopic factors (GSF's) $\F_{N0}$ and $\F_{N1}$
for  hypernuclei up to $^{29}_\Lambda $Si we perform  summations  over $\lambda$ in \rf{2.16}
 for each single-particle state $j=p_ {3/2},p_ {1/2}$, and  $j=d_ {5/2}$.
The resulting  $\R_{NJ}^{j}$ turn out to be  quite similar to \rf{2.24} for
$\R_{NJ}^{s_ {1/2}}$, except that now $P_{N}$, and
$\O_0(P)$ are substituted, respectively, by  $P^{j}_{N}$, and  $\O_{l_N}(P)$.
Thus, we supplement   the plain  $\sWA$ with the substitution,
 \br
P^j_{N}&\go& P_{N},
\nn\\
\O_{l_N}(P)&\go&\O_0(P),
 \label{2.28}\er
which are  fair approximations for the evaluations of ratios $\Gamma_n/\Gamma_p$ and $a_\Lambda$.
\footnote{This $\sWA$ has been used in Ref.~\cite{Ba07} to relate
 the matrix elements $\M(plPL\lambda SJ\T;j_\Lambda j_N=0p_{3/2}, J\tLN)$, and
 $(plSJ\T|V|{\sf l}=0,JJ\tLN)$ in $_\Lambda^{12}$C  (see ~\cite[Eq. (B2)]{Ba07}.
There are two misprints in~\cite[Eq. (B2)]{Ba07}. The correct results are:
 ${\cal M}(p2,P1,1110;\Lambda \p)=\frac{1}{2\sqrt{6}}{\sf  d}(p)(P1|11)$, and
${\cal M}(p2,P1,2120;\Lambda \p)=\frac{\sqrt{3}}{2\sqrt{10}}{\sf  d}(p)(P1|11)$.}
In this way we get
\br
\R_{N1}^{p_ {3/2}}&=&
\frac{R_{N0}}{3}+\frac{R_{N1}}{6},
\hspace{0.5cm}
\R_{N2}^{p_ {3/2}}=\frac{R_{N1}}{2},
\nn\\
\R_{N1}^{p_ {1/2}}&=&
\frac{R_{N0}}{6}+\frac{R_{N1}}{3},
\hspace{0.5cm}
\R_{N0}^{p_ {1/2}}=\frac{R_{N1}}{2},
\nn\\
\R_{N2}^{d_ {5/2}}&=&
\frac{3R_{N0}}{20}+\frac{R_{N1}}{10},
\hspace{0.5cm}
\R_{N3}^{d_ {5/2}}=\frac{R_{N1}}{4}.
\label{2.29}\er
where the numerical factors come from the summation on $\lambda$
of the squares of coefficients
$C_0(l\lambda SJl_Nj_N;K)$ given by \rf{2.22}.

It could be useful to  express the Eq. \rf{2.24}
within the Block-Dalitz notation~\cite{Bl63}:
 \begin{equation}\label{2.30}
\begin{array}[b]{lll}
{\sf a} = \bra{ ^1\!\mathrm{S}_0}{V}\ket{ ^1\!\mathrm{S}_0},
\quad
&
{\sf b} = \bra{ ^3\!\mathrm{P}_0}{V}\ket{ ^1\!\mathrm{S}_0},
\quad
&
{\sf c} = \bra{ ^3\!\mathrm{S}_1}{V}\ket{ ^3\!\mathrm{S}_1},
\\
& & \\
{\sf d} = \bra{ ^3\!\mathrm{D}_1}{V}\ket{ ^3\!\mathrm{S}_1},
\quad
&
{\sf e}= \bra{ ^1\!\mathrm{P}_1}{V}\ket{ ^3\!\mathrm{S}_1},
\quad
&
{\sf f} = \bra{ ^3\!\mathrm{P}_1}{V}\ket{ ^3\!\mathrm{S}_1},
\end{array}
\end{equation}
 for the NME. Assuming the same value of single-particle energies
 for protons and neutrons in the state $0s_{1/2}$, one gets the well known results
\br
R_{n0}&=&2({ a}^2+{ b}^2),\hspace{1cm}R_{p0}={a}^2+{b }^2,
\nn\\
R_{n1}&=&2{f}^2,\hspace{1cm}R_{p1}={c}^2+{ d}^2+{ e}^2+{f}^2,
\label{2.31}\er
where
\br
&&{ a}=
\frac{M_{\rm N}}{\pi}\sqrt{\frac{A}{A-2}}\nn\\
&\x&\int^{P_N }_0 dPP^2\sqrt{P_N^2-P^2}\O_0^2(P){\sf a}^2(p),
\label{2.32}\er
and similarly for ${ b},\cdots{ f}$.
As the NME depend very weakly on the momentum $p$ we can  compute them   at $p=
p_\Delta\equiv\sqrt{M_{\rm N}\Delta}$~\cite{Ba07}, and write
\br
&&{ a}=\J_0
{\sf a}^2(p_\Delta),\etc
\label{2.33}\er
with
\br
&&\J_0=
\frac{2M_{\rm N}}{\pi}\sqrt{\frac{A}{A-2}}\nn\\
&\x&\int^{P_N }_0 dPP^2\sqrt{P_N^2-P^2}\O_0^2(P),
\label{2.34}\er
which after performing the integration reads~\cite{Gr65}
\br \J_0
&=&2M_{\rm N}^2\sqrt{2\pi \frac{A-2}{A}}\Delta_{N}b e^{-z}I_{1}(z),
\label{2.35}\er
where $I_{1}(z)$ is modified Bessel function of the first kind, and
\br
z=\Delta_{N}M_{\rm N}b^2\left(\frac{A-2}{A}\right).
\label{2.36}\er
Using the  asymptotic form of $I_1(z)$
($I_1(z)\cong{e^z}/{\sqrt{2\pi z}}$)
 one gets
\br \J_0
=2\sqrt{ M_{\rm N}^3\Delta_{N}}\cong 2M_Np_\Delta,
\label{2.37}\er
which is the same result as that  derived previously in Ref.~\cite {Ba07},
where the recoil effect  was not  considered. Therefore one sees that
after performing the integration in \rf{2.32} this effect  is totally
 washed out from  the transition rates.

To evaluate the $\Gamma_N$ from \rf{2.14}, as well as to derive the GSF's in \rf{2.27}, we need to
know the spectroscopic factors $F^j_{NJ}$ given by \rf{2.15}. These, in turn, depend on
the angular momenta $J_C$ and $J_I$, which are fixed from the experimental data, and are
exhibited in Table \ref{table1}.
The $F^j_{NJ}$-values
 for most of the  hypernuclei discussed here
are listed in~\cite[Table I]{Kr03}. The remaining can be easily inferred from this
 table,  except for the $p$-wave ones in $^7_\Lambda $Li
 ($F^{p_{3/2}}_{N1}=5/8$, and $F^{p_{3/2}}_{N2}=3/8$), and in
$^9_\Lambda $Be ($F^{p_{3/2}}_{N1}=3/4$, and $F^{p_{3/2}}_{N2}=5/4$).
The resulting GSF's $\F_{NJ}$  are
listed in Table  \ref{table1}.
 It is noticeable that in the $jj$ closed shells the
singlet to triplet ratio $(\F_{N0}:\F_{N1})$ is always $(1:3)$.

\begin{table}[h] \caption{ Core spins $J_C$, initial  spins $J_I$, and the generalized
spectroscopic factors $\F_{NJ}$ within the $jj$-coupling.
 } \label{table1}
\bigskip
\begin{tabular}{|c|c|cccc|}
\hline
$^A_\Lambda Z$&$J_C,J_I$  &$\F_{n0}$&$\F_{n1}$&$\F_{p0}$&$\F_{p1}$\\
\hline
$^5_\Lambda $He &$0,\fot$&$1/2$&$3/2$&$1/2 $&$3/2$\\
 $^7_\Lambda $Li &$1,\fot$&$17/24$&$43/24$&$17/24$&$43/24$\\
 $^{9}_\Lambda $He& $0,\fot$
&$1 $&$3$&$1/2$&$3/2$\\
 $^9_\Lambda $Be &$0,\fot$&$5/8$&$15/8$&$5/8$&$15/8$\\
 $^{9}_\Lambda $C& $0,\fot$
&$1/2$&$3/2$&$1 $&$3$\\
  $^{11}_\Lambda$B&$3,\frac{5}{2}$
&$25/24$&$59/24$&$25/24$&$59/24$\\
  $^{12}_\Lambda $C& $\frac{3}{2},1$
 &$13/12$&$29/12$&$1 $&$3$\\
  $^{13}_\Lambda $C&$0,\fot$
&$1$&$3$&$1 $&$3$\\
  $^{21}_\Lambda $C&$0,\fot$
&$13/8$&$39/8$&$1 $&$3$\\
  $^{16}_\Lambda $O&$\frac{1}{2},1$
 &$7/6$&$10/3$&$5/4 $&$15/4$\\
  $^{17}_\Lambda $O& $0,\fot$
   &$5/4 $&$15/4$&$5/4 $&$15/4$\\
  $^{28}_\Lambda $Si&$\frac{5}{2},2$
  &$33/20$&$23/5$&$13/8 $&$39/8$\\
       $^{29}_\Lambda $Si&$0,\fot$
&$13/8 $&$39/8$&$13/8 $&$39/8$\\
  \hline
   \end{tabular}
\end{table}

In all hypernuclei heavier than $^5_\Lambda $He the contribution of the state $0s_{1/2}$
to the total rate $\Gamma_{N}$ is
\br
\Gamma_{N}^s&=&\frac{1}{2}R_{N0}+\frac{3}{2}R_{N1},
\label{2.38}\er
while the contributions  of the single-particle
states  $0p_{3/2}$, $0p_{1/2}$, etc depend on their occupations,
which, in turn,  are reflected in the values of the GSF's listed in
Table \ref{table1}. For instance, $\Gamma_{p}^{p_{3/2}}=
\Gamma_{p}^{s}$, and $\Gamma_{n}^{p_{3/2}}= \Gamma_{n}^{s}$
 in all hypernuclei with ${\sf Z}\ge 6$, and ${\sf N}\ge 6$, respectively.
  In the same way, $\Gamma_{p}^{p_{1/2}}= \Gamma_{p}^{s}/2$,
and $\Gamma_{n}^{p_{1/2}}= \Gamma_{n}^{s}/2$
 in all hypernuclei with ${\sf Z}\ge 8$, and ${\sf N}\ge 8$, respectively.
The orbital $0d_{5/2}$ supplies less transition strength than the $0s_{1/2}$ state,
and it is given by
\br
\Gamma_{N}^{d_{5/2}}&=&\frac{3}{8}\left(R_{N0}+3R_{N1}\right),
\label{2.39}\er
in all hypernuclei with ${\sf Z}\ge 14$, or ${\sf N}\ge 14$.

Several years ago, by means of  the  $\sWA$,  Cohen~\cite{Co90}
arrived at the estimate
\br
\Gamma_{n/p}&=&\frac{R_{n0}+3R_{n1}}{R_{p0}+3R_{p1}},
\label{2.40}\er
 for the "heavy species" of hypernuclei.
From Table \ref{table1} one  sees, however, that this relation is:
i)  strictly fulfilled only
for ${\sf N}={\sf Z}$ nuclei, \ie $^5_\Lambda $He,  $^{9}_\Lambda $Be, $^{13}_\Lambda $C,
$^{17}_\Lambda $O, and $^{29}_\Lambda $Si, ii)  approximately correct
for hypernuclei with ${\sf N}\cong {\sf Z}$, and iii)
 totally invalid for hypernuclei far from the stability line.

Moreover, as  all hypernuclei with the same
 $A$ hold the same  elementary rates \rf{2.24}, very simple relationships
 can be established from Table \ref{table1}
between their rates $\Gamma_p$, and $\Gamma_n$, and   ratios $\Gamma_{n/p}$.
 For instance:
\br
\Gamma_{p}(^{9}_\Lambda{\rm He})&=&0.8\Gamma_{p}(^{9}_\Lambda{\rm Be})
=0.5\Gamma_{p}(^{9}_\Lambda{\rm C}),
\nn\\
\Gamma_{n}(^{9}_\Lambda{\rm He})&=&1.6\Gamma_{n}(^{9}_\Lambda{\rm Be})
=2\Gamma_{n}(^{9}_\Lambda{\rm C}),
\nn\\
\Gamma_{n/p}(^{9}_\Lambda{\rm He})&=&2\Gamma_{n/p}(^{9}_\Lambda{\rm Be})
=4\Gamma_{n/p}(^{9}_\Lambda{\rm C}).
\label{2.41}\er
These and similar results for other values of $A$ are very likely
 independent of both the decay mechanism
and the final state interactions.

Detailed calculations of Refs.~\cite{Pa97,It02,Ba02} have proved that  the contribution
of the $p$ partial wave to NMWD in $p$-shell hypernuclei, as well as in
heavy-mass systems,   is relatively
small ($\lsim 10 \%$). In particular,
Itonaga \etal~\cite{It02} have explored  the decay rates $\Gamma_n$, and $\Gamma_p$ in hypernuclei
from $A=4$ up to $A={209}$, establishing that the $p$-wave contributions
to the calculated total one-nucleon induced decay rates $\Gamma_{nm}=\Gamma_{p}+\Gamma_{n}$,
and ratios $\Gamma_{n/p}$ are only a few percent
of the respective $s$-wave contributions (see~\cite[Fig. 9]{It02}).
They have attributed this finding  to the
short range of the decay interaction.
In fact, it is well known that the ranges of the radial pieces
 $v(r)$ of the  OME potentials are inversely proportional to the meson masses
(see~\cite[Figs. 3, 4, and 5]{It02}), being the largest inverse
mass  that of the pion ($m_\pi^{-1}=1.4$ fm). Then, when one
analyzes the radial matrix element  in \rf{2.19}, which reads
(see~\cite[Eq. (A19)]{Ba03}) \br (pl|v(r)|0{\sf l})&=&\int
r^2drj_l(pr)g_{NN}v(r)g_{N\Lambda}{\rm R}_{0{\sf l}}(\sqrt{2}b,r),
\nn\\
\label{2.42}\er
where $g_{N\Lambda}$ and $g_{NN}$ are, respectively, the initial and final SRCs functions,
one can see that the harmonic oscillator wave function ${\rm R}_{0{\sf l}}(\sqrt{2}b,r)$
is: i) picked at the origin for ${\sf l}=0$,  and ii)
 the integrand maxima  for ${\sf l}=1$, which in principle should be
 at the distance $\sqrt{2}b$ ($=2.5$ fm for $^{12}_\Lambda $C) from the origin,
 is shifted  even farther because of the factor $r^2$.
 This, together with
the approximation \rf{2.28} for the c.m. overlaps,   makes
the $s$-wave radial matrix elements
large  compared  to the $p$-wave ones. Moreover,
it could be worth mentioning  that with the Wood-Saxon radial  wave functions
one gets  analogous  results since they are quite similar to
that  of the harmonic oscillator,  as can be seen, for instance, from
~\cite[Figure 2-22]{Bo69}.

\section{Comparison between exact and ${\rm\bf s}$WA results }
 For  the hypernuclei of interest here, the approximated results are confronted
   numerically  with the full calculations in Table \ref{table2}.
 This is done within the following framework: a) The
NMWD dynamics is  described  by the $\pi+K$  OME potential, with
the weak coupling constants from Ref.~\cite{Pa97,Pa02}, b) The
parameter $b$ is evaluated as in Ref.~\cite{Kr03}, \ie
$b=1/\sqrt{\hbar \omega M_{\rm N}}$, with $\hbar \omega
=45A^{-1/3}-25A^{-2/3}$ MeV.
 c) The  initial and final SRCs,  as well as  the finite nucleon size effects are included
in the same way as in our previous works~\cite{Ba07,Ba02,Ba03,Kr03}. The results  displayed in Table
\ref{table2} clearly show that the agreement between the exact and
$\sWA$ results is indeed quite satisfactory. In fact, the differences
between them are of the same order of magnitude or smaller than that of the kinematical
and nonlocality effects discussed in~\cite{Ba03}.

 \begin{table}[h]
\caption{ Results for exact,  and
 $\sWA$ transition rates, evaluated, respectively, from   Eqs. \rf{2.14}, and \rf{2.27}.
 Those for $^{9}_\Lambda $Be are
not shown, since  they fall in between  those for $^{9}_\Lambda $He,
and $^{9}_\Lambda $C.}
\label{table2}
\bigskip
\begin{tabular}{|cccc cc|}
\hline
$^A_\Lambda Z$ &Approx.&$\Gamma_n$&$\Gamma_p$&$\Gamma_{nm}$&$\Gamma_{n/p}$\\
\hline $^5_\Lambda $He &Exact&
 $0.149$&$   0.358$&$   0.507$&$0.417$\\
 $^7_\Lambda $Li &Exact&
$   0.154$&$   0.375$&$   0.529$&$   0.409$\\
 &$\sWA$ &$   0.153$&$   0.369$&$   0.523$&$   0.416$\\
 $^{9}_\Lambda $He&Exact&$   0.265$&$   0.317$&$   0.583$&$   0.836$\\
&$\sWA$ &$   0.262$&$   0.318$&$   0.581$&$   0.824$\\
 $^{9}_\Lambda $C&Exact&$   0.131$&$   0.676$&$   0.807$&$   0.194$\\
&$\sWA$ &$   0.131$&$   0.637$&$   0.768$&$   0.206$\\
  $^{11}_\Lambda$B&Exact&
$   0.208$&$   0.528$&$   0.736$&$   0.394$\\
&$\sWA$ &$   0.207$&$   0.499$&$   0.706$&$   0.414$\\
  $^{12}_\Lambda $C&
Exact&$   0.200$&$   0.628$&$   0.828$&$   0.319$\\
&$\sWA$ &$   0.199$&$   0.594$&$   0.794$&$   0.335$\\

 $^{12}_\Lambda $C' &Exact&$0.205$&$0.793$&$0.998$&$0.259$\\
 &$\sWA$        &$0.201$&$0.755$&$0.955$&$0.266$\\

  $^{13}_\Lambda $C&Exact&$   0.241$&$   0.615$&$   0.855$&$   0.391$\\
&$\sWA$ &$   0.238$&$   0.582$&$   0.820$&$   0.409$\\
  $^{21}_\Lambda $C&Exact&$0.340  $&$     0.535  $&$     0.874   $&$    0.635$\\
&$\sWA$ &$   0.335$&$   0.510$&$   0.845$&$   0.657$\\
  $^{16}_\Lambda $O&Exact&$   0.253$&$   0.719$&$   0.972$&$   0.352$\\
  &$\sWA$ &$   0.250$&$   0.689$&$   0.938$&$   0.362$\\
  $^{17}_\Lambda $O&Exact&$   0.279$&$   0.710$&$   0.989$&$   0.393$\\
  &$\sWA$ &$   0.275$&$   0.677$&$   0.952$&$   0.406$\\
  $^{28}_\Lambda $Si &Exact&$   0.297$&$   0.815$&$   1.112$&$   0.364$\\
  & $\sWA$ &$   0.289$&$   0.760$&$   1.049$&$   0.380$\\
      $^{29}_\Lambda $Si&Exact&$   0.311$&$   0.806$&$   1.117$&$   0.385$\\
&$\sWA$ &$   0.302$&$   0.751$&$   1.053$&$   0.401$\\   \hline
   \end{tabular}
\end{table}

 A more rigorous inclusion of  the strong interaction ingredients on the initial and  final two-body states,
 as done in Refs.~\cite{Pa02,It02,It08}, will   modify  in the same way the exact and $\sWA$ results, without affecting
 the conclusions of the present work.  Namely, there is no physical reason why
  the  mixing between states with the same total angular momenta and   different orbital and spin angular
 momenta - induced by the SRCs, and exhibited  in ~\cite[Eqs. (46) and (47)]{It08} - should influence
differently the exact and $\sWA$ calculations.
That this is true for the final state follows immediately from
 the fact that the  $\sWA$ is done only on the initial state.
Thus, all the discussion performed in  Refs.~\cite{Pa02,It08} for the final-state tensor correlation
 is equally valid for
both calculations. The initial $\Lambda N$ SRCs  are less discussed in the literature. Nevertheless,
 it was established that the phenomenological spin-independent correlation
 function~\cite[Eq. (21)]{Pa97}, which is the same as that  used
 here, is a good approximation of the full correlation function.

The $\sWA$ works very well for any other choice of the OME potential different from
the model assumption of  $\pi+K$  exchanges considered above. As one example
in Table \ref{table2} are also shown  the results for
$^{12}_\Lambda $C (labelled as $^{12}_\Lambda $C')
obtained with
the full  $\pi+ \eta+ K+\rho+\omega+K^*$ OME  potential.
It is also evident that more realistic estimates
of the oscillator parameter $b$, as the one given by Itonaga \etal~\cite{It02},
would affect both calculations in the same manner, and would yield to the same degree of
agreement
between the exact and $\sWA$ results \footnote {The  $b$-values used here:
$1.765,1.781,1.838$, and $1.966$ fm , for $^{11}_\Lambda $B,
 $^{12}_\Lambda $C,  $^{16}_\Lambda $O, and  $^{28}_\Lambda $Si, do not differ much from
 the values reported in Ref.~\cite{It02}, which are, respectively, $1.65,1.65,1.755$, and $1.865$.}.

\begin{figure}[h]
\includegraphics[width=8.6cm,height=9.0cm]{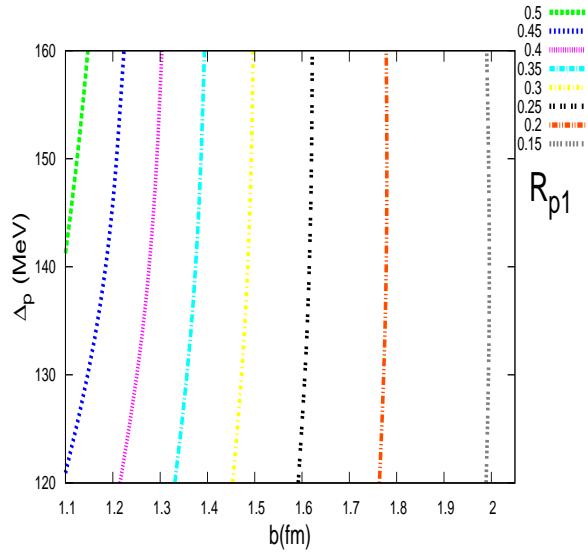}
\caption{\label{Figure1} (Color online) Single particle decay rate
$R_{p1}$ as a function of the length parameter $b$ and the
liberated energy $\Delta_p$.}
\end{figure}

One should keep in mind that in the Fermi gas model the
$\Lambda$-hyperon is taken to
 be always in a relative
$s$-state with respect to any of the nucleons within the
hypernucleus. Therefore,
 the success of the $\sWA$ indirectly justifies the application  of such a model to the NMWD
 of finite nuclei~\cite{Bau07}.

After fixing the OME potential, all $R_{NJ}$ depend only on $b$ and $\Delta_N$. As an example,
the dependence   of
 $R_{p1}$ on these two quantities is illustrated in Fig. \ref{Figure1}. The variation of
$\Delta_p$ has a very small effect, as can be seen from Eq. \rf{2.37}.
 Contrary to this, the  $R_{NJ}$   depend very strongly on $b$ trough
the radial wave function ${\rm R}_{00}(\sqrt{2}b,r)$ in Eq. \rf{2.42}.
The corresponding $s$-shell single-particle decay rates  $R_{NJ}$  are exhibited
in Table  \ref{table3}. Finally, we note that
by using the values listed
 in Tables \ref{table1} and \ref{table3},
together with  Eq. \rf{2.27},
we recover the $\sWA$ results shown in Table \ref{table2}.

\begin{table}[h]
\caption{The s-shell single
particle decay rates $R_{NJ}$ scaled by a factor of 10.
The results for $^{9}_\Lambda $C, and $^{9}_\Lambda $Be
are not shown as they are the same as those for $^{9}_\Lambda $He.}
\label{table3}
\bigskip
\begin{tabular}{|c|ccc cc|}
\hline
$^A_\Lambda Z$ &&$10R_{n0}$&$10R_{n1}$&$10R_{p0}$&$10R_{p1}$\\
\hline
$^5_\Lambda $He &&
$       0.1411$&$       0.9470$&$       0.0705$&$       2.3620$\\
 $^7_\Lambda $Li &&
$       0.1310$&$       0.8842$&$       0.0655$&$       2.2194$\\
$^{9}_\Lambda $He&&
$       0.1228$&$       0.8334$&$       0.0614$&$       2.1026$\\
 $^{11}_\Lambda$B&&
$       0.1162$&$       0.7915$&$       0.0581$&$       2.0054$\\
 $^{12}_\Lambda $C&&
$       0.1133$&$       0.7732$&$       0.0567$&$       1.9626$\\
  $^{13}_\Lambda $C&&
$       0.1107$&$       0.7563$&$       0.0553$&$       1.9230$\\
 $^{21}_\Lambda $C&&
$       0.0950$&$       0.6557$&$       0.0475$&$       1.6845$\\
$^{16}_\Lambda $O&&
$       0.1038$&$       0.7124$&$       0.0519$&$       1.8196$\\
 $^{17}_\Lambda $O&&
$       0.1018$&$       0.6997$&$       0.0509$&$       1.7894$\\
  $^{28}_\Lambda $Si&&
$       0.0860$&$       0.5975$&$       0.0430$&$       1.5441$\\
       $^{29}_\Lambda $Si&&
$       0.0849$&$       0.5905$&$       0.0425$&$       1.5272$\\
  \hline
   \end{tabular}
\end{table}

Table \ref{table3}, as well as  Fig. \ref{Figure1},
 clearly show that the size parameter $b$ is the most important nuclear structure
 parameter for the NMWD  rates $\Gamma_{n}$ and  $\Gamma_{p}$,
 and therefore the knowledge of its value for each individual
 hypernuclei could become  crucial in comparing the theory with  experiments.
 However, this not come to pass with the ratio $\Gamma_{n/p}$, which is mainly tailored by the
OME potential.

\section{ Conclusions and Summary}

The following conclusions can be drawn regardless of the  OME potential that is used:

(1) The $\sWA$  is sufficiently accurate, not only for qualitative
discussions, but also for quantitative descriptions of the NMWD in
hypernuclei within the IPSM, when  the SRCs are described by
phenomenological correlation functions as done here.

(2) The increase of  transition rates $\Gamma_{n}$, $\Gamma_{p}$, and $\Gamma_{nm}$, as a
 function of the hypernuclear mass number, stems from the interplay of the increase
  of   $\F_{NJ}$, and the decrease of   $R_{NJ}$, and

(3) The ratio  $\Gamma_{n/p}$ is almost the same for
all hypernuclei that are on the stability line (${\sf N}={\sf Z}$), \ie $^5_\Lambda $He, $^7_\Lambda $Li,
$^{11}_\Lambda$B,  $^{13}_\Lambda $C,
 $^{17}_\Lambda $O, $^{29}_\Lambda $Si, \etc. Moreover,
it decreases  when one moves toward the proton drip-line (${\sf Z}>{\sf N}$),
and increases when one goes toward the neutron drip-line (${\sf N}>{\sf Z}$).
It diminishes, for instance, by more than a factor of 4 when going from  $^9_\Lambda $He
to  $^9_\Lambda $C, while  the constituent $R_{NJ}$ rates remain the same.
It might be somewhat surprising that $\Gamma_{n}<\Gamma_{p}$ even
when  the neutron number is
greater than the proton number. But, as seen from Table \ref{table3}, the reason for this  is
the dominance of $\R_{p1}$ on the other three single-particle decay rates.  This dominance,
in turn, comes from the dominance of the tensor amplitude $d$ on the remaining amplitudes.
The only exception is  $^4_\Lambda $H for which $\R_{p1}$ does not contribute.

In summary, using as a tool the IPSM and the $s$-wave approximation, we have
shown that  the decay  rates $\Gamma_{n}$, and $\Gamma_{p}$
 can be interrelated in a very simple way in all hypernuclei
going from $^5_\Lambda{\rm He}$ up to $^{29}_\Lambda $Si.
The relationships between them are particularly simple for the hypernuclei with the same
mass number, as illustrated by  Eq.  \rf{2.41} for the sequence
$^9_\Lambda $He $\go ^9_\Lambda $Be$\go ^9_\Lambda $C.
Results of this type are very likely valid in general,
 and as such they  could be exploited to study experimentally
 the variations of $\Gamma_{n}$,  $\Gamma_{p}$ and  $\Gamma_{n/p}$ along many
similar  arrays in a systematic way.

\begin{acknowledgments}
This work is supported by the Argentinian agency CONICET under
contract PIP 0377. I am  grateful to Eduardo
Bauer for  helpful discussion and critical reading of the manuscript.
\end{acknowledgments}

\end{document}